\documentclass[aaspp4]{aastex}

\def\deg{$^\circ$}

\slugcomment{To appear in \apj}

\begin{document}

\title{Nuclear spirals as feeding channels to the Supermassive 
Black Hole: the case of the galaxy NGC\,6951}

\author{Thaisa Storchi-Bergmann, Oli L. Dors Jr. \& Rogemar A. Riffel}
\affil{Instituto de F\'\i sica, Universidade Federal do Rio Grande do Sul, Av. Bento
Gon\c calves 9500, 91501-970 Porto Alegre RS, Brazil}

\author{Kambiz Fathi}
\affil{Instituto de Astrof\'\i sica de Canarias, C/ V\'\i a L\'actea s/n, 38200 La Laguna, 
Tenerife, Spain }

\author{David J. Axon}
\affil{Physics Department, Rochester Institute of Technology, 
85 Lomb Memorial Drive, Rochester, New York 14623, USA}

\author{Andrew Robinson }
\affil{Physics Department, Rochester Institute of Technology, 
85 Lomb Memorial Drive, Rochester, New York 14623, USA}

\author{Alessandro Marconi}
\affil{INAF - Osservatorio Astrofisico di Arcetri,
Largo Fermi 5, I-50125 Firenze, Italy}

\author{G\"oran \"Ostlin}
\affil{Stockholm Observatory, AlbaNova University Center, 106 91 Stockholm, Sweden}


\begin{abstract}

We report the discovery of gas streaming motions along nuclear
spiral arms towards the LINER nucleus of the galaxy NGC\,6951.
The observations, obtained
using the GMOS integral field spectrograph on the Gemini North
telescope, yielded maps of the flux distributions and gas kinematics 
in the H$\alpha$, [N{\sc\,ii}]$\lambda$6584 and 
[S{\sc\,ii}]$\lambda\lambda$6717,31 
emission lines of the inner 7\arcsec$\times$15\arcsec\ of the galaxy.
This region includes a circumnuclear star-forming ring with radius $\sim$500\,pc,
a nuclear spiral inside the ring and the LINER nucleus. 
The kinematics of the ionized gas is dominated by rotation, but
subtraction of a kinematic model of a rotating exponential disk reveals
deviations from circular rotation within the nuclear ring
which can be attributed to (1) streaming motions along the nuclear 
spiral arms and (2) a bipolar outflow which seems to be
associated to a nuclear jet.
On the basis of the observed streaming velocities and geometry
of the spiral arms we estimate a mass inflow rate of ionized gas
of $\approx\,3\times\,10^{-4}$M$_\odot$\,yr$^{-1}$, which is
of the order of the accretion rate necessary to power the
LINER nucleus of NGC\,6951. Similar streaming motions towards 
the nucleus of another galaxy with LINER nucleus -- NGC\,1097 -- 
have been reported by our group in a previous paper. Taken together,
these results support a scenario in which nuclear 
spirals are channels through which matter is transferred 
from galactic scales to the nuclear region to feed the
supermassive black hole.


\end{abstract}


\section{Introduction}

One long-standing problem in the study of nuclear activity in
galaxies is to understand how mass is transferred from galactic
scales down to nuclear scales to feed the supermassive black hole
(hereafter SMBH) inside.  Many theoretical studies 
and simulations (Shlosman et al. 1990; Emsellem et al. 2003; Knapen 2005;
Emsellem et al. 2006) have shown that non-axisymmetric 
potentials efficiently promote gas
inflow towards the inner regions (Englmaier \& Shlosman 2004).
Recent observations have  revealed that
structures such as small-scale disks or nuclear bars and associated
spiral arms are frequently observed in the inner kiloparsec of active
galaxies (Erwin \& Sparke 1999; Pogge \& Martini 2002;
Laine et al. 2003).

In a recent work, Lopes et al. (2007)
have shown a strong correlation between the presence of nuclear
dust structures (filaments, spirals and disks) and activity
in galaxies. Nuclear spirals, in particular, 
are estimated to reside in more 
than half of active galaxies (Martini et al. 2003). 
Martini \& Pogge (1999) have shown that nuclear spirals are not 
self-gravitating, and that they are likely to be shocks in nuclear 
gas disks. Simulations by Maciejewski (2004a,b) demonstrated that, if a central SMBH 
is present, spiral shocks can extend all the way to the SMBH vicinity and 
generate gas inflow consistent with the accretion rates inferred in 
local AGN.

The above studies support the
hypothesis that nuclear spirals are a mechanism for fueling
the nuclear SMBH, transporting gas from  kiloparsec
scales down to a few tens of parsecs of the active
nucleus (Knapen et al. 2000; Emsellem et al. 2001;
Maciejewski et al. 2002; Marconi et al. 2003;
Crenshaw et al. 2003; Fathi et al. 2005).
This hypothesis has recently been confirmed by our 
group (Fathi et al. 2006) in the specific case of 
the LINER/Seyfert 1 galaxy NGC\,1097.
Using Integral Field Spectroscopy at the Gemini telescope 
we mapped the velocity field of the ionized gas and  
detected streaming motions towards the nucleus
along nuclear spiral arms.
 
NGC\,1097 is the only case so far in which
streaming motions along nuclear spiral arms have been mapped. 
In order to verify if such spirals are always associated with inward
streaming motions it is necessary to map the
gas kinematics in more galaxies. With this
goal in mind, we have obtained Gemini Integral Field 
spectroscopic observations of a few more active galaxies 
with nuclear spirals observed in HST images. 
The sample was selected from nearby galaxies (z$<$0.005)  
with intermediate inclinations, to facilitate study of the gas  
kinematics on linear distance scales of tens of parsecs.

In the present paper we report the results for 
NGC\,6951, a galaxy with Hubble type SAB(rs)bc, 
at a distance of 24\,Mpc (Tully 1988), such that 1 arcsec 
corresponds to 96\,pc at the galaxy. 
Originally it was argued that NGC\,6951 has a LINER type nucleus 
(Filippenko \& Sargent 1985), but 
more recently it has been suggested that actually its activity 
is intermediate between LINER and Seyfert 
(P\'erez et al. 2000). NGC\,6951 hosts a large 
scale bar (with total extent of $\sim$4\,kpc), and 
at about 5\arcsec\ (480\,pc) from the nucleus, there is a
conspicuous star-forming ring, previously observed in H$\alpha$
(M\'arquez \& Moles 1993; P\'erez et al. 2000) and showing
also strong CO and HCN emission (Kohno et al. 1999,
Garcia-Burillo et al. 2005, Krips et al. 2007). Continuum radio emission
has also been observed from the nucleus and star-forming
ring  by Saikia et al. (1994), while a higher resolution radio 
image can be found in Ho \& Ulvestad (2001).

Our IFU measurements show that, similarly to NGC\,1097, 
the gas kinematics inside the nuclear ring of NGC\,6951, 
although dominated by circular rotation, shows deviations 
in the measured radial velocities which can be interpreted
as at least partially due to streaming inward motions along nuclear spiral arms.
Our results strenghten the  case that nuclear spirals
are indeed channels to feed the supermassive black hole in 
active galaxies.

The present paper is organized as follows. In Section 2
we describe the observations and reductions. In Section 3
we present the flux and kinematic measurements. In Section 4
we discuss the results and in Section 5 we present
our conclusions.

\section{Observations and Reductions}

The observations were obtained with the Integral Field Unit of
the Gemini Multi-Object Spectrograph (GMOS-IFU, Allington-Smith et 
al. 2002) at the Gemini North telescope, on
the nights of August 31 and September 1, 2006. The observations
consisted of three adjacent IFU fields
(covering $ 5 \times 7 \: \rm arcsec^{2}$ each)  resulting in a total
angular coverage of $ 7 \times 15 \: \rm arcsec^{2}$ around 
the nucleus. Three  exposures  of  500\,s were obtained for each
of the 3 IFU fields, slightly shifted in order to correct 
for detector defects. 
Observations of the three IFU fields were  
obtained consecutively on the same night. The fluxes in adjacent IFU  
pixels from neighbouring fields were found to be consistent within 10\%. 
Therefore, no shifts or scaling were applied when combining the  
three fields to obtain the final mosaic.

The longest extent of the three adjacent IFU fields was
oriented along position angle (hereafter PA) 140$^\circ$,
selected to approximately coincide with the major axis of the galaxy,
(PA=138\deg\ according to M\'arquez \& Moles 1993), while
the large scale bar is oriented approximately along E--W.
Each IFU fiber has a diameter which corresponds to 
0.2\,\arcsec\ in the sky, while the seeing
during the observations ranged between 0.4 and 0.5\,\arcsec, 
corresponding to a spatial resolution at the galaxy of $\approx$\,40\,pc.

The selected wavelength range was 5600--7000\AA,
in order to cover the H$\alpha$+[N{\sc\,ii}]$\lambda\lambda$6548,84 
and [S{\sc\,ii}]$\lambda\lambda$6716,31
emission lines, observed with the grating GMOS R400-G5325 (set to central
wavelength $\approx$ 6300 \AA), with an instrumental FWHM of
2.9\AA, corresponding to a spectral resolution 
of R$\approx 2\,300$ ($\approx 130$\,km\,s$^{-1}$).

The data reduction was performed using  
the specific tasks developed for GMOS data in the 
{\sc gemini.gmos} package as well as generic tasks in {\sc iraf}. 
The reduction process comprised
bias subtraction, flat-fielding, trimming, wavelength calibration,
sky subtraction, relative flux calibration, building of the data cubes
at a sampling of 0.1$\arcsec$\,$\times$\,0.1$\arcsec$,
and finally the alignment and combination of the 9 data cubes.
As we have obtained only relative flux calibration (what is called
in Gemini ``baseline calibrations''), we have normalized the flux
levels using the H$\alpha$ nuclear flux
reported by Perez et al. (2000).

\section{Results}

\begin{figure*}
\centering
\epsscale{1}
\plotone{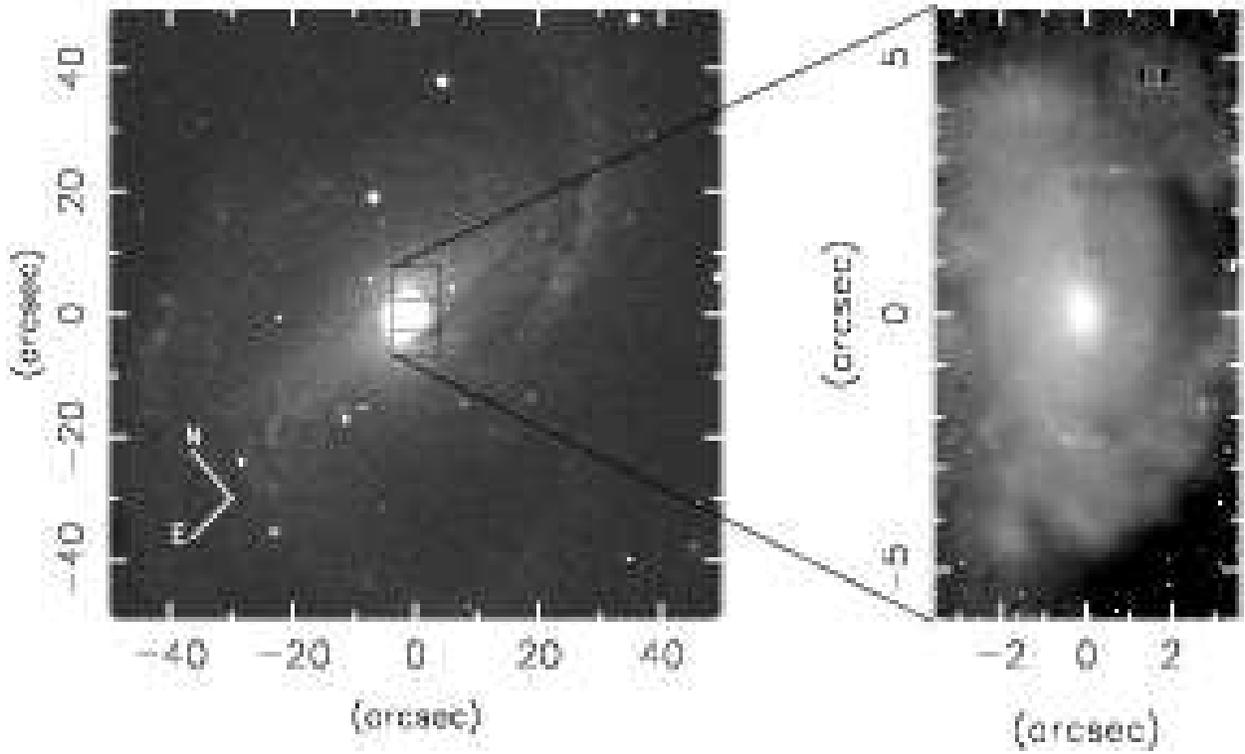}
\caption{Left: The acquisition image in the R-band showing 
the three IFU fields covered by the observations. Right: Image 
obtained from the IFU spectra in the continuum adjacent to the H$\alpha$+[N{\sc\,ii}] 
emission lines. The sampling of the data is 0.1\arcsec$\times$0.1\arcsec, and the seeing was $\approx$0.4\arcsec.}
\label{image}
\end{figure*}

In Figure~\ref{image} we show the acquisition image with
the location of the three IFU fields, together with 
an image obtained from the IFU data in the continuum
adjacent to the H$\alpha$ emission-line.
The IFU observations cover the nuclear region out to the star-forming ring,
more clearly seen in the H$\alpha$ and [N{\sc\,ii}] images shown in 
Fig.~\ref{flux_ratios}. The H$\alpha$ and [N{\sc\,ii}] emission lines
could be measured over most of the $ 7 \times 15 \: \rm arcsec^{2}$ field,
except for a few locations which appear in black in Fig.~\ref{flux_ratios}. For the
[S{\sc\,ii}] emission lines, there are more locations where the lines could not be measured due to the low signal-to-noise ratio in these lines. 

\clearpage
\clearpage

\begin{figure*}
\plotone{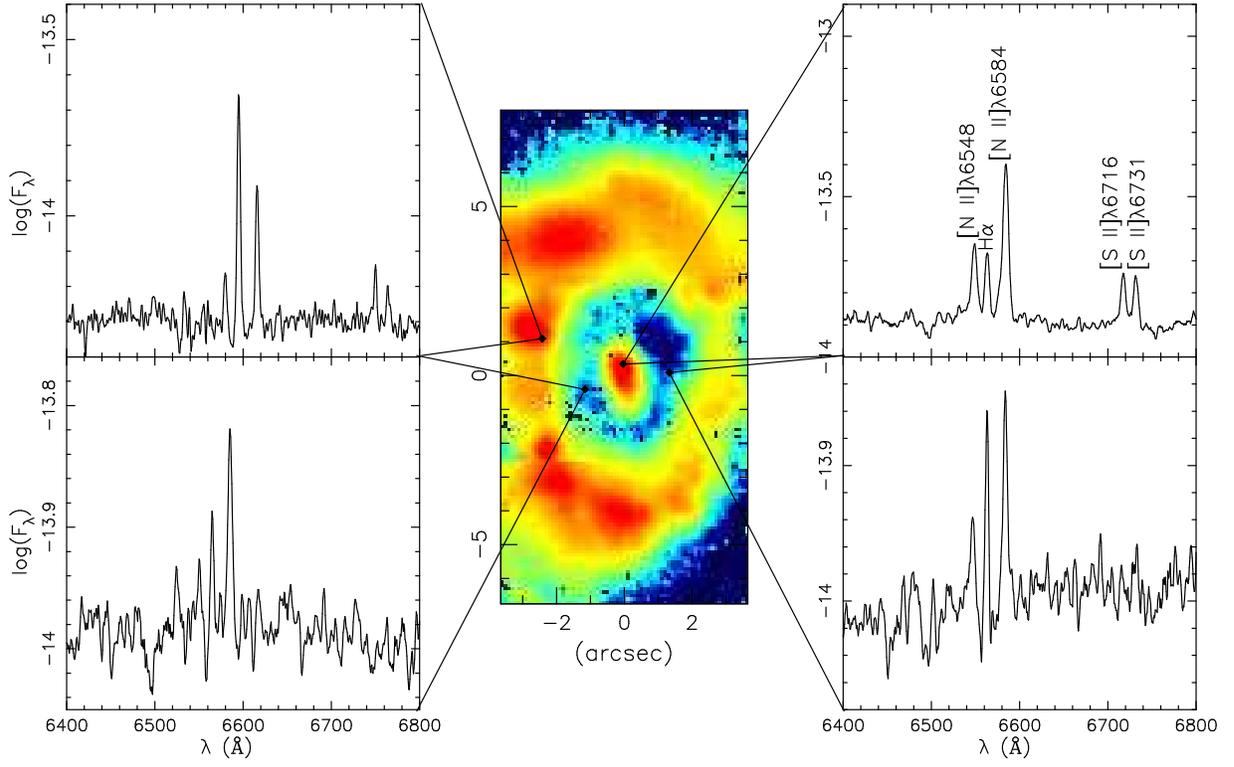}
\caption{Sample of spectra, with the corresponding locations identified in
the IFU image obtained from the integrated flux in the H$\alpha$ 
emission line (central panel). Upper right: the nuclear spectrum;
upper left: a spectrum from the star-forming ring; bottom left and right:
spectra from regions between the nucleus and star-forming ring.}
\label{spectra}
\end{figure*}

In Figure~\ref{spectra} we show a sample of spectra: one from the
nucleus, one from  the star-forming ring and two from locations between 
the nucleus and the ring. The emission-lines present in most spectra are H$\alpha$, 
[N{\sc\,ii}]$\lambda\lambda$6548,84 and
[S{\sc\,ii}]$\lambda\lambda$6717,31. 
The lines [O{\sc\,i}]$\lambda\lambda$6300,63
were  marginally detected only in a few spectra, and were not considered 
in our analysis due to the poor signal-to-noise ratio (in these lines).
The emission-line spectrum from the nucleus is LINER-like ([N{\sc\,ii}]$\lambda$6584
stronger than H$\alpha$), as already known from previous works (e.g. Filippenko \& Sargent 1985). More recently, Ho, Fillipenko \& Sargent (1997) have 
argued that the nuclear activity is better classified as Seyfert 2,
after an analysis of the nuclear spectrum 
taking into account the effect of the underlying stellar absorption
in the fluxes of the Balmer lines. P\'erez et al. (2000),
after a similar analysis, conclude that the nuclear activity of NGC\,6951 
is intermediate between LINER and Seyfert 2. Our data are consistent with
all these classifications, as our restricted wavelength coverage does
not allow the observation of additional line ratios necessary to
distinguish between the different types of activity.

The emission-line ratios in the star-forming ring
are typical of H{\sc\,ii} regions, while
in the regions between the nucleus and the ring, the line-ratios are
also mostly LINER-like, turning to ``H{\sc\,ii}-like'' as the star-forming ring
is approached.
\clearpage
\clearpage

\subsection{Line fluxes and excitation of the emitting gas}

\begin{figure*}
\epsscale{1}
\plotone{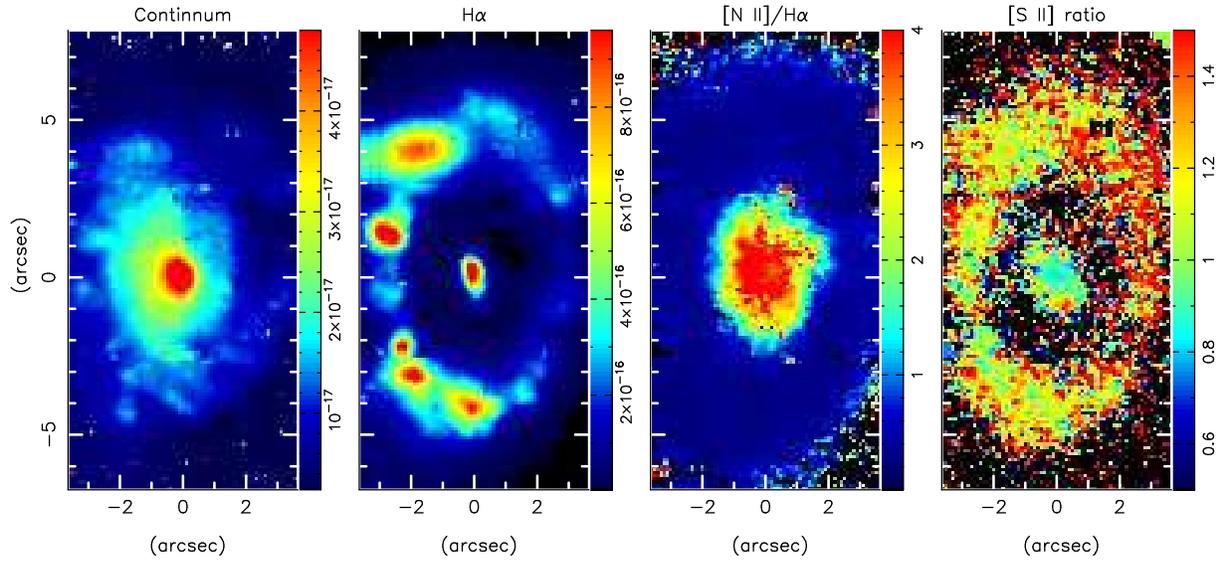}
\caption{From left to right: map of the continuum flux adjacent to H$\alpha$
(in units of erg\,cm$^{-2}$\,s$^{-1}$\,\AA$^{-1}$ per pixel); 
map of the H$\alpha$ integrated flux (erg\,cm$^{-2}$\,s$^{-1}$ per pixel); 
line-ratio maps [N{\sc\,ii}]/H$\alpha$ and [S{\sc\,ii}]6717/6731.}
\label{flux_ratios}
\end{figure*}

The line fluxes, radial velocities and velocity dispersions were obtained 
by fitting Gaussians to the emission-lines, which accurately
represent the observed line profiles. The H$\alpha$ image,
obtained from the integrated H$\alpha$ flux is shown in 
Fig.~\ref{flux_ratios}, together with the image in the
adjacent continuum. Most spectra have signal-to-noise ratios 
in the H$\alpha$ and [N{\sc\,ii}] emission lines larger than 50.

The star-forming ring is clearly seen in the H$\alpha$ image, 
formed by several H{\sc\,ii} regions which appear as emission-line knots
along the ring. The ring is more extended along the major axis of the galaxy
(PA$\sim$140\deg), where it reaches $\approx$ 4.5\arcsec\ from the nucleus
(430\,pc at the galaxy),
while in the perpendicular direction it reaches only $\approx$ 2.5\arcsec.

The line ratio map [N{\sc\,ii}]/H$\alpha$, obtained from the ratio
of the images in [N{\sc\,ii}]$\lambda$6584
and H$\alpha$, as well as the [S{\sc\,ii}]$\lambda$6717/$\lambda$6731
ratio map are also shown in Fig.~\ref{flux_ratios}.
Prior to the construction of the map [N{\sc\,ii}]/H$\alpha$
we have corrected the H$\alpha$ line fluxes by an average
underlying absorption due to the stellar population 
with an equivalenth width of 0.75\AA.
This procedure is necessary in order to avoid unphysical very high ratios
of [N{\sc\,ii}]/H$\alpha$ and is supported by the inspection of a few
spectra which reveal that where H$\alpha$ emission is very faint
the line is indeed obviously filling an absorption. The average value of
0.75\AA\ was adopted from the work of P\'erez et al. (2000)
who estimated an absorption equivalent width of H$\alpha$ of 1\AA\ at
the nucleus and 0.5\AA\ at the ring on the basis of the analysis of the
stellar population. We thus decided to adopt an average value 
between that of the nucleus and that of the ring, as the resulting
corrected fluxes did not change by more than $\sim$20\% at the nucleus and
$\sim$10\% at the ring if we change the underlying absorption
by 0.25\AA.

The highest [N{\sc\,ii}]/H$\alpha$ line ratios reach values of 4-5 
within a $\approx$1\arcsec\ radius from the nucleus, decreasing to 2-3 between
1\arcsec\ and 2\arcsec\ and down to $\approx$0.3 at the circumnuclear ring.

The ratio [S{\sc\,ii}]$\lambda$6717/$\lambda$6731 reaches a minimum value
of $\approx$0.8, implying a maximum gas density of $\approx\,10^3$\,cm$^{-3}$ within 
$\approx$\,1\arcsec\ from the nucleus (Osterbrock 1989). At the 
star-forming ring, the [S{\sc\,ii}]
ratio ranges from 1 to 1.4, with typical values at most locations between 
1.1 and 1.2, implying gas densities in the range 250--400\,cm$^{-3}$.
At the borders of the ring, the low-density limit value of 1.4 ($<$100\,cm$^{-3}$)
is reached. In most locations between the nucleus and the ring, 
the signal-to-noise ratio was too low to get a reliable value for the [S{\sc\,ii}]
ratio. In many location where it could be measured, it was in the low-density
limit. In order to investigate this further, we have binned 9 spectra
together (which corresponds to 0.3\,$\times$\,0.3\,arcsec$^2$, 
and thus approximately to the angular resolution of the data) 
in several locations and got ratios in the range 1.2$<$[S{\sc\,ii}]6717/6731$<$1.4, thus somewhat above the low-density limit,
implying ionized gas densities of $\sim$100\,cm$^{-3}$, which we will
adopt as the typical value for this region.

\clearpage
\clearpage

\subsection{Kinematics}

\begin{figure*}
\epsscale{0.60}
\plotone{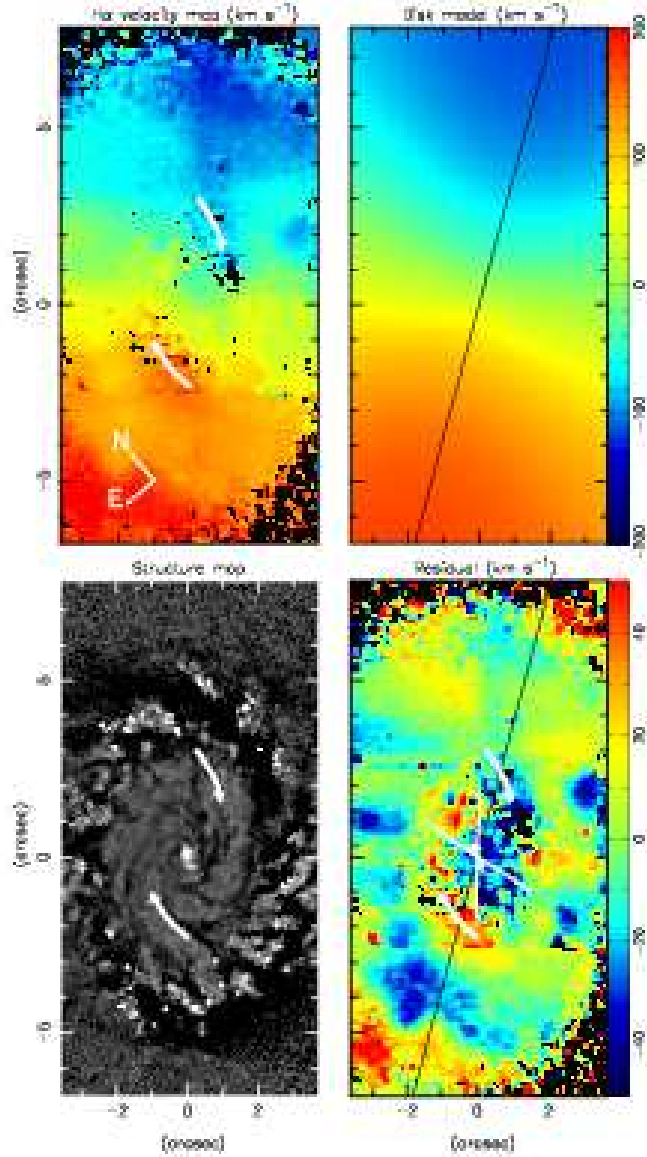}
\caption{Top left: the H$\alpha$ velocity field. Top right:
model velocity field. Bottom left: structure map.
Bottom Right: difference between observed and model velocity fields.
Curved lines with arrows identify two velocity structures 
which deviate from the model: the ``W arm''\,(top) and 
the ``E arm''\,(bottom). The small bar with arrows at the
nucleus in the bottom right panel represents a compact radio source (Saikia et al. 2002),
and the two white lines delineate a bi-cone within which we observe
an enhancement in the values of $\sigma$ (see Fig. \ref{sigma}).
The straight black line shows the location of the line of nodes.} 
\label{radial_vel}
\end{figure*}

The central wavelengths of the Gaussian curves fitted to the emission-lines
were used to obtain the gas radial velocities, while the full widths at half
maxima (FWHM) were used to obtain the gas velocity dispersions 
($\sigma=$FWHM/2.355). The accuracy of the velocity measurements is $\approx$10\,km\,s$^{-1}$. The radial velocity
field obtained from H$\alpha$ is shown in the top left panel of
Fig.~\ref{radial_vel}.  It is dominated 
by rotation, as evidenced by its similarity to the classical ``spider diagram''
(Binney \& Merrifield 1998). (The pixels for which we could not measure the
kinematics due to low signal-to-noise ratio in the
emission lines are shown in black in the figure.)

There are, nevertheless, clear deviations from simple rotation in the
radial velocity field. In order to isolate these deviations, 
we have fitted an exponential thin disk kinematic model to the
H$\alpha$ radial velocity data,
assuming that the density profile is exponential and its kinematics is 
circular rotation (Freeman 1970; Binney \& Tremaine 1987). 
The procedure is similar to the one we have adopted in modelling the
circumnuclear gas kinematics of 
NGC\,1097 (Fathi et al. 2005; Fathi et al. 2006).
 
The velocity field corresponding to the  best fit is 
shown in the top right panel of Fig.~\ref{radial_vel}. The fit to the model 
gives a systemic velocity of 1450$\pm$20\,km\,s$^{-1}$, a deprojected 
maximum velocity amplitude of 220$\pm$10\,km\,s$^{-1}$, a disk scale length of 
4.1\arcsec$\pm$0.3\,(390\,pc at the galaxy) and 
a position angle for the line of nodes of PA=125\deg$\pm$10\deg. We have adopted 
a disk inclination of $i=42$\deg, a value derived from photometry
(M\'arquez \& Moles 1993), because the inclination 
is not well constrained  in our multi-parameter space fitting procedure.

\clearpage
\clearpage

A comparison between the measured radial velocity field and the 
disk model reveals deviations from simple rotation.
These deviations are observed along the star-forming ring
as well as inside the ring, where, in particular, 
we have identified two features resembling partial spiral arms
marked by two white curved lines with arrows 
in Fig.~\ref{radial_vel}. In order to investigate 
if these deviations correspond to real features in the galaxy, we have
constructed a structure map (Pogge \& Martini 2002) using an HST image 
of the nuclear region of NGC\,6951 obtained through the filter F606W
(HST proposal 8597, PI M. Regan).
The structure map is useful to enhance the contrast in the images, 
thus showing more clearly both dark structures -- regions obscured by dust,
as well as bright regions such as H{\sc\,ii} regions, a compact bright nucleus or star clusters.
The structure map is shown in the bottom left panel of Fig.~\ref{radial_vel},
where we have drawn the same white lines of the top left panel, showing that
the  partial spiral arms follow the dusty spiral structure
seen in the structure map. Although there are more spiral arms 
in the structure map than the two we see in the velocity residual map,
this difference can be understood as due to the
poorer angular resolution of the IFU data ($\approx$0.4\arcsec)
compared to that of the  HST (0.05\arcsec), 
which precludes us from establishing a more detailed correspondence.
In the bottom right panel of Fig.~\ref{radial_vel} we show the residuals between
the observed and the model velocity field, where the
white lines correspond again to the partial arm structures. 
The top arm (hereafter called W arm)
corresponds to a blueshifted region in the residual map, while the bottom
arm (hereafter E arm) corresponds to a redshifted region. 
There are aditional blue- and  redshift residuals in the star forming ring
as well as close to the nucleus. The aditional velocity residuals 
close to the nucleus in Fig.~\ref{radial_vel} are approximately delimited by
the borders of a ``bi-cone'', sketched in white in the figure, which corresponds
to regions in which we see an increase in the values of the velocity dispersion,
as discussed below.

\begin{figure*} 
\epsscale{0.8}
\plotone{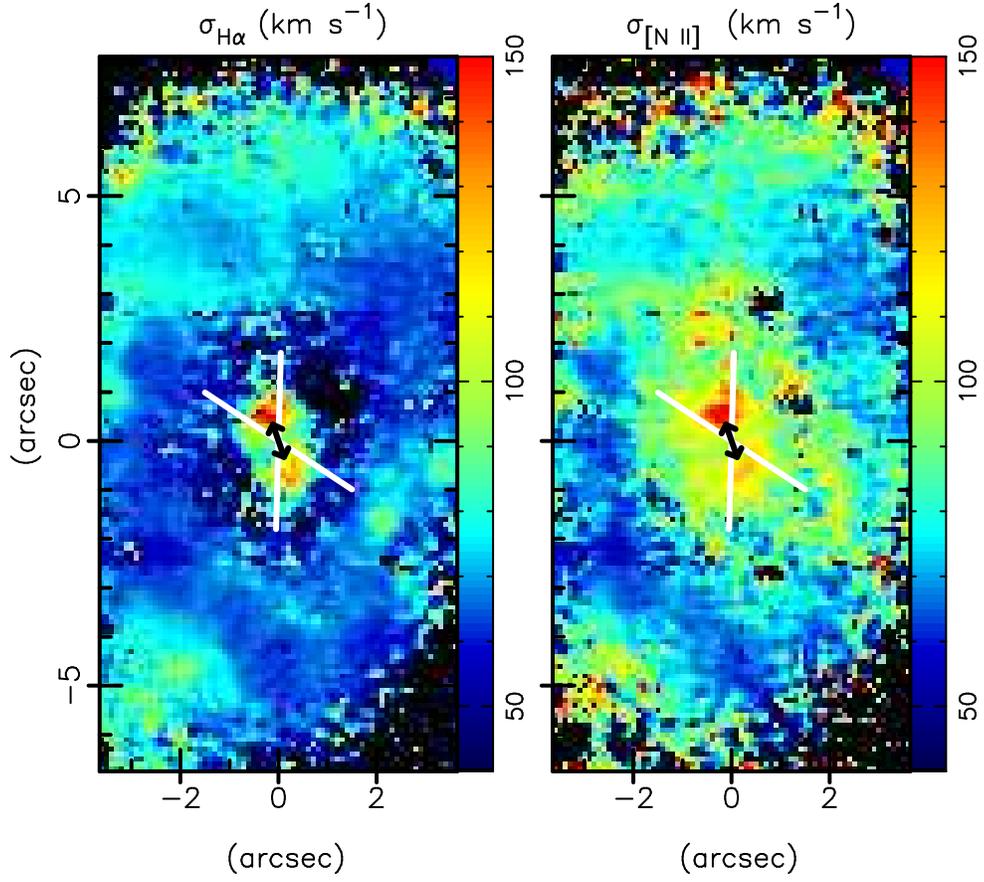}
\caption{Velocity dispersion maps in H$\alpha$ (left) and [N{\sc\,ii}]$\lambda$6584 (right).
The black bar with two arrows represents the compact nuclear radio source (Saikia et al. 2002), while the two white lines delineate the borders of a bi-cone, whithin which the highest $\sigma$ values are observed, probably due to stirring of the circumnuclear ISM by a nuclear radio jet.}
\label{sigma}
\end{figure*}

The gas velocity dispersion ($\sigma$) maps in H$\alpha$ and [N{\sc\,ii}] 
are shown in Fig.\,\ref{sigma}.
The instrumental $\sigma$ is $\approx$50\,km\,s$^{-1}$, thus it can be concluded that
many values, particularly for H$\alpha$, are at the resolution limit of the instrument. 
The low values obtained for H$\alpha$ in some regions are due to the fact that, in the
region between the nucleus and the star-forming ring H$\alpha$ emission is 
strongly affected by the underlying stellar absorption (discussed above), 
which makes the observed emission line weaker and narrower.
On the other hand, the [N{\sc\,ii}] profile provides a reliable measure 
of $\sigma$ because it is free of the contamination by stellar absorption.
Most $\sigma$ values range between 60 and 80\,km\,s$^{-1}$ but there are larger
values reaching 140\,km\,s$^{-1}$ in two blobs within $\approx$1\arcsec\ from
the nucleus to the S-SE and N-NW. 

\clearpage
\clearpage

What could be the origin of these enhancements in $\sigma$?
Our experience with previous IFU observations (of other targets; e.g. Riffel
et al. 2006; Barbosa et al. 2007), reveal that increases in 
$\sigma$ in the gas surrounding AGN are usually related to the presence
of a passing radio jet. We searched the literature for a possible 
nuclear radio jet and found that VLA observations by Saikia et al. (2002) 
of the inner region of NGC\,6951 indeed suggest the presence
of a resolved nuclear radio source presenting an
angular size of 0.7$\times$0.2\,arcsec$^2$ along PA=156\deg. We show a
representation of this small radio source as a bar with 
arrows in Fig.\,\ref{sigma} as well as
in the bottom right panel of Fig.\,\ref{radial_vel}. Notice that the
extent and orientation of this radio structure coincide with those of a bright similarly elongated nuclear structure seen in the structure map (bottom left panel of Fig.\,\ref{radial_vel}). 
As the F606W filter used to obtain the HST image includes H$\alpha$,
this coincidence suggests that we are resolving enhanced H$\alpha$ 
emission associated with the radio structure in this image.

It can be seen that the estimated extent of the radio source fits well in between 
the two regions of enhanced $\sigma$ values in Fig.\,\ref{sigma}.
We have sketched (in white in the figure) the borders of a bi-cone
which includes the regions of largest $\sigma$ values. We notice that
the corresponding locations in Fig.~\ref{radial_vel} approximately
delimitate also a redshifted H$\alpha$ emitting region 
to the N--NW and a blue-shifted region to the S--SE
observed in the residual H$\alpha$ velocity map (bottom right panel).
The observed increase in $\sigma$ along the bi-cone and
the presence of associated blue- and reshifts suggests that these two 
structures are due to an outflow driven by the radio jet. The delineated 
bi-cone in Figs. \ref{radial_vel} and \ref{sigma} approximately separates
the regions affected by the outflow from the regions affected by the
residual motion along the spiral arms.

\section{Discussion}

\subsection{Excitation}

An extensive photometric and spectroscopic study of the galaxy NGC\,6951 has been presented 
by P\'erez et al. (2000, hereafter P00). They adopt the disk major axis PA=138\deg\ and 
have obtained longslit spectra (slit width of 1 arcsecond) along this PA as well as along PA=48\deg\ (minor axis) and PA=84\deg.

P00 quote line ratio values of [N{\sc\,ii}]/H$\alpha$ larger than unity within the inner 2\arcsec,
reaching $\approx$5 at the nucleus, in agreement with our results. They present data within the nuclear region at a sampling of 0.5\arcsec\ (although the seeing was 1.2\arcsec\ during their observations). P00 report electron density values
increasing from N$_e\sim\,$300\,cm$^{-3}$ at the ring to $\sim\,$1000\,cm$^{-3}$ at the nucleus, also in agreement with our results. 

\begin{figure*} 
\plotone{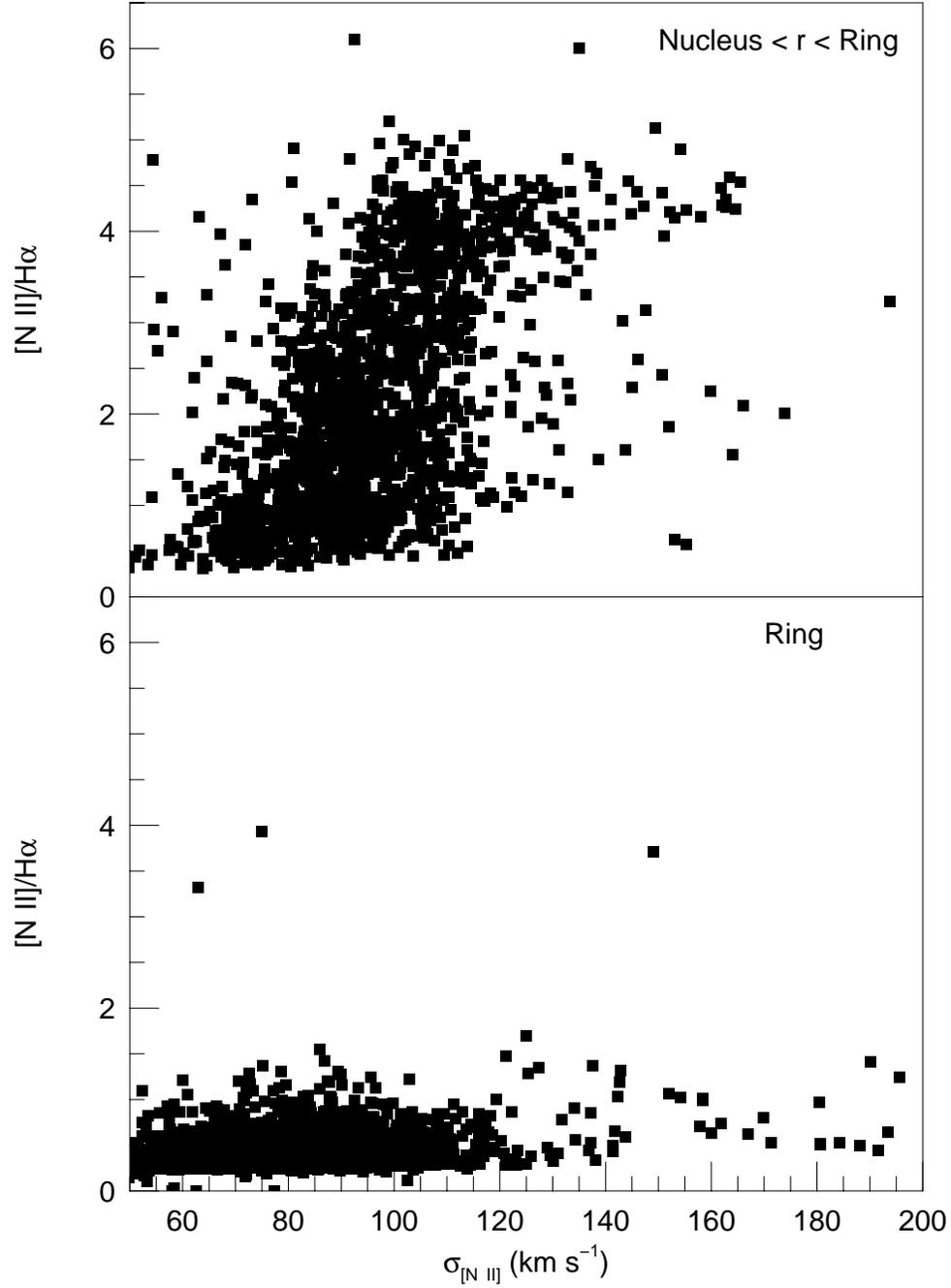}
\caption{The ratio [N{\sc\,ii}]$\lambda$6584/H$\alpha$ as a function of the [N{\sc\,ii}]$\lambda$6584 velocity dispersion for the region inside the ring (top)
and for the ring (bottom).} 
\label{excitation}
\end{figure*}

In order to further investigate the excitation, we have plotted in Fig.\,\ref{excitation}
the ratio [N{\sc\,ii}]/H$\alpha$ vs. the velocity dispersion obtained from the width of the
[N{\sc\,ii}]$\lambda$6584 emission-line $\sigma_{[N{\sc\,ii}]}$. 
We have separated the data from the ring and from the region inside the ring
using the H$\alpha$ image to identify the region dominated by the ring. 
In order to do this, we have fitted two ellipses: one to the inner border of the
ring and another to the outer border. Along the longest extent of the
ring (approximately along the vertical axis of the figures), the inner
border is located at 2.8\arcsec\ from the nucleus and the outer border
at 6.8\arcsec, while along the shortest extent of the ring
the inner border is at 2.0\arcsec\ and the outer border at 3.6\arcsec.

It can be seen that the behavior of the two regions
is clearly different: while for the ring most values of [N{\sc\,ii}]/H$\alpha$ are 
in the range 0.3--0.4, typical of H{\sc\,ii} regions and show just a weak correlation 
with $\sigma_{[N{\sc\,ii}]}$,
in the region inside the ring the [N{\sc\,ii}]/H$\alpha$ values are much higher 
and present a strong correlation with $\sigma_{[N{\sc\,ii}]}$. This behavior has 
been observed previously in other galaxies (e.g. Keppel et al. 1991;
Sokolowski et al. 1991; Storchi-Bergmann et al. 1996), 
meaning that the gas excitation is related
to its kinematics, and suggest in particular that at least part of the
ionization is produced by shocks as shocks do enhance the [N{\sc\,ii}]/H$\alpha$
ratio (Viegas \& Contini 1994; Sutherland et al. 1993), as well as 
broaden the emission-lines. The presence of shocks in the vicinity
of the nucleus is particularly evident in Fig.\,\ref{sigma}, where we
have shown that two regions which have the highest observed $\sigma$ values 
are located at the ends of the nuclear radio structure, and can be interpreted
as gas from the interstellar medium of the galaxy shocked by 
a radio jet.
\clearpage
\clearpage
\subsection{The nuclear spiral}

Garcia-Burillo et al. (2005, hereafter GB05) have obtained
molecular CO radio images of the inner region of NGC\,6951, showing that
most of the CO emission is concentrated in ``two winding nuclear spiral arms
which end up as a pseudo-ring at $\sim$\,350\,pc from the nucleus'',
with an estimated molecular gas mass of $\approx\,3\times10^8$\,M$_\odot$.
This nuclear spiral seen in molecular gas corresponds in fact to the
H$\alpha$ ring, whose strongest H$\alpha$ emission seems to be a bit external to the
strongest molecular gas emission.

GB05 also report the observation of ``an unresolved nuclear complex''
within 50\,pc from the nucleus, estimating its mass to be a few $\sim10^6$\,M$_\odot$.
The presence of a relatively large molecular gas mass at the nucleus is supported also
by the recent mm-interferometric observations by Krips et al. (2007)
who have found compact HCN emission at the nucleus and estimate an even 
larger mass for the corresponding molecular complex, of $\sim$\,2--10$\times\,10^7\,$M$_\odot$.
These observations support the 
recent inflow of molecular gas to the nuclear region. 

Besides the nuclear complex, GB05 have also ``tentatively detected''
(their words) low level CO emission  bridging the gap between
the nuclear source and the N part of the ring, with an estimated molecular mass of 
$\sim\,10^7$\,M$_\odot$. GB05 argue that this structure could be the molecular
counterpart of the filamentary spiral structure identified in a
J--H color map of NGC\,6951 obtained from HST images. Their J--H map (Fig. 4a in GB05) 
is very similar to the structure map we show
in the lower left panel of Fig.\,\ref{radial_vel}. A comparison of this
structure map with both our H$\alpha$ velocity map (upper left panel of Fig.\, \ref{radial_vel})
and Fig.\,4 of GB05 shows that their ``molecular bridge'' 
seems to partially coincide with the W arm structure seen in our
kinematic map, in particular with the ``inner end'' of our partial W arm.
A close look shows that the molecular bridge 
seems to be associated with the darker regions of the structure map closer to the
nucleus (an thus with the dust structure) while our W partial arm 
is better coincident with the brighter portions of the structure map, a bit
more distant from the nucleus.
A possible interpretation is that we are seeing two parts of a spiral arm:
a more external one, traced by H$\alpha$ emission, and a more
internal one, seen only in molecular gas emission, probably due to the 
absence -- or faintness -- of ionized gas emission.

\subsection{Kinematics}

The rotation pattern in the H$\alpha$ radial velocity field 
(Fig.\,\ref{radial_vel}) shows that the SE side of the galaxy is receding 
while the NW side is approaching. 
Under the assumption that the spiral arms are trailing, and that the
approaching side of the bar piles up the dust in front of it, 
it can be concluded 
that the near side of the galaxy is the SW, and the far side is the NE.
The deviations from circular rotation along the two partial spiral 
arms in the inner 2--3\arcsec\ from the nucleus -- blueshifted 
residuals along the W arm and redshifted 
residuals along the E arm -- can thus be interpreted as
streaming motions along nuclear spiral arms towards the nucleus.

Nevertheless, the residuals in this inner region extend beyond 
the region covered by the E and W 
arms drawn in Fig.\,\ref{radial_vel}: in particular, there
are additional blueshifts observed
to the S--SE of the nucleus and redshifts observed to the N--NW,
which seem to be a continuation of the E and W arms. 
Nevertheless, at $\sim$1\arcsec N--NW and S--SE of the
nucleus, they are not consistent 
with streaming motions towards the
nucleus.  We conclude that these residuals are in fact 
outflows due to the interaction
of the radio source at the nucleus (Saikia et al. 2002) 
with the circumnuclear ISM of the galaxy. 

This hypothesis is supported by the increase in the 
velocity dispersion of the gas from 70\,km\,s$^{-1}$ to 140\,km\,s$^{-1}$ 
along the position angle of the radio structure (PA=156\deg), 
approximately within a region delineated by two white 
lines representing the borders of a bi-cone in Fig.\,\ref{sigma}. 
From the location of this 
bi-cone in Fig.\,\ref{radial_vel}, we conclude that  
the residual blueshifts at $\approx$1\arcsec N-NW
and redshifts at $\approx$1\arcsec S-SE of the nucleus are
observed in the regions corresponding to those with enhanced $\sigma$
and can be interpreted as outflows related to the radio source.

The residuals from circular motion in the inner 3\arcsec\ thus
include two components: (1) streaming motions towards the nucleus
along spiral arms and (2) an outflow due to the interaction of 
a radio source with the circumnuclear ISM of the galaxy. From the
relative intensity of the continuum observed in the structure map
in the location of the radio structure,
which is brighter to the side to which the residuals are blueshifts,
we suggest that the near side of the jet is projected against the 
near side of the galaxy, which is possible only if its angle relative
to the galaxy plane is smaller than 48\deg, for the adopted inclination
of the galaxy of 42\deg.

In the regions corresponding to the ring, there are also
radial velocity residuals relative to the circular rotation
which could be due to outflows from the starbursts along the ring
and/or deviations due to the accumulation of the gas in the
Inner Lindblad Ressonance. 

The kinematics we have obtained for the nuclear region of NGC\,6951 
can be compared with that
obtained in previous works, in particular by P00, 
who have studied the large scale kinematics from long-slit spectroscopy 
along three position angles (over a radius of up to 100\arcsec\ from the 
nucleus): one along the photometric major axis (138\deg), 
one along the perpendicular direction (48\deg) and one along PA=84\deg. 
On the basis of the kinematics along these three PAs, P00 
obtain a systemic velocity of $V_{sys}=1417\pm$\,4\,km\,s$^{-1}$ and an 
amplitude of the rotation curve of $\approx$\,200\,km\,s${-1}$, after
de-projection (see, for example, Fig. 13 of P00). These parameters
are similar to our values of $V_{sys}=1450\pm$\,20\,km\,s$^{-1}$ and 
amplitude $220\pm10$\,km\,s$^{-1}$ of the rotation curve. 
These paramenters are also similar to those previously
obtained by M\'arques \& Moles (1993) on the basis of
longslit spectra along the major and minor axes of the galaxy, in spite
of the fact that our kinematic data is two-dimensional but covers a much
smaller radial extent (3.5\arcsec\,$\times\,7\arcsec$). This difference in 
angular coverage may explain the difference between the major axis orientation obtained
in previous studies and the one derived from our kinematic modelling
of 125$\pm$\,10\,\deg.  

P00 report a ``peculiar structure'' in the velocity field along the
minor axis (PA=48\deg), which they argue could be due to a nuclear
outflow or to a structure related to the ``internal 
Inner Lindblad Ressonance (iILR)''. They adopt the second 
hyppothesis, suggesting counter rotation 
of an inner disk, which they argue could be related to the iILR, 
calculated to occur at 180\,pc (2\arcsec). 
They argue that this would fit into a scenario of disk nested within disks.
Our data, with a complete 2D coverage, favors best
the nuclear outflow hypothesis, but in combination 
with streaming motions along nuclear spirals.

\subsection{The origin of the nuclear spiral}

GB05 have used near-IR HST images of the nuclear region -- where they found
a stellar oval --  in combination with the CO observations, in order to estimate whether 
torques produced by the non-axisymmetric stellar potential would be efficient enough to drain the gas angular momentum allowing its inflow to the nuclear region. GB05 conclude
that the inflow is efficient only down to the star forming ring, approximately. 
Inside the ring, the torques are positive and cannot drive gas inflow. As molecular gas
is observed much closer to the nucleus, GB05 conclude that, inside the ring,
other mechanisms should be responsible to drive the AGN fueling, such as viscous torques. 


Englmaier \& Shlosman (2000) have used 2D numerical simulations to show that
nuclear gaseous spirals are generated as as result of the gas response to the 
gravitational potential of a galactic disk with a large scale stellar bar.
Both NGC\,6951 and NGC\,1097 (Fathi et al. 2006) present similar large 
scale stellar bars, which could thus play a role in the formation of the
nuclear spirals. 

More recently, Maciejewski (2004) has considered the effect of a SMBH in 
nuclear gaseous spirals.
The fact that the SMBH does play a role is supported by the ubiquitous presence
of nuclear spirals in Seyfert galaxies (Martini \& Pogge 1999; Pogge \& Martini 2002),
and, in particular, by the strong correlation of their presence with the
nuclear activity (Lopes et al. 2007). Maciejewski concludes that,
if the asymmetry is strong, spiral shocks are produced, and the central
SMBH can allow the spiral shocks to extend all the way to its immediate vicinity,
generating gas inflow of up to 0.03\,M$_\odot$\,yr$^{-1}$.
The cases of NGC\,6951 presented here and NGC\,1097 presented by Fathi et al. (2006) seem
to support the models of Maciejewski (2004a,b).

\subsection{Estimating the mass accretion flow}

We can use the residual velocities observed along the nuclear spiral 
to estimate the rate of mass flowing to the
nucleus. In order to do that, we assume that the mass flows
along two nuclear spiral arms. The cross section of these arms can be 
estimated assuming an opening angle of 20\deg\ subtended by 
the spiral arms at 100\,pc from the nucleus. 
This opening angle corresponds to a radius for an assumed
circular cross section of the spiral arm calculated as: 

\begin{equation}
r \approx {\rm 100\,(pc)\ \frac{10\,\pi}{180}= 17.5\,(pc)}
\end{equation}
\\
The flux of matter crossing the circular cross section can be calculated as:
\begin{equation}
\Phi= N_e\ v\,\pi\,r^2\,n_{arms}\,m_p\,f
\end{equation}
\noindent
where $N_e$ is the electron density, $v$ is the streaming velocity
of the gas towards the nucleus, $n_{arms}$ is the number of spiral
arms, $m_p$ is the mass of the proton, and $f$ is the filling factor.
The filling factor was estimated from:

\begin{equation}
 L_{H\alpha} \sim f\,N_e^2\,j_{H\alpha}(T)\,V
\end{equation}
\noindent
where $L_{H\alpha}$ is the H$\alpha$ luminosity emitted by a volume $V$ and
$j_{H\alpha(T)}=$3.534$\times10^{-25}$\,erg\,cm$^{-3}$\,s$^{-1}$
(Osterbrock 1989). In the region covered by the partial spiral arms we obtain an average
flux per pixel (of angular size 0.1\arcsec$\times$\,0.1\arcsec) of 4.8$\times10^{-17}$\,erg\,cm$^{-2}$\,s$^{-1}$.
In order to calculate the H$\alpha$ luminosity in the region covered by a
partial arm, we adopt an area of 2\arcsec\,$\times$1\arcsec\, which corresponds to 
200 pixels. We thus obtain $L_{H\alpha}$\,=\,200$\,\times$\,4.8$\,\times10^{-17}$\,erg\,cm$^{-2}$\,s$^{-1}
\times\,4\pi\,(24$\,Mpc)$^2$=6.62$\times$\,10$^{38}$\,erg\,s$^{-1}$.
Adopting an approximate volume for the region (from Fig.\,\ref{radial_vel})
of 2\arcsec\,$\times$1\arcsec\,$\times\,$1\arcsec, or $V$\,=\,2000$\,\times$\,(9.6\,pc)$^3$,
we obtain $f\approx$\,0.004.

The velocity residuals in the region of 
the partial nuclear arms, which we have interpreted as streaming motions,
range from $\approx$\,20\,km\,s$^{-1}$ to $\approx$\,50\,km\,s$^{-1}$ (Fig.\,\ref{radial_vel}). Assuming that these streaming motions occur in 
the plane of the galaxy, we need to correct these values for the inclination of the galaxy
$i=42$\deg. The resulting average velocity for the streaming motions is $\approx v$\,=\,40\,km\,s$^{-1}$. For $N_e=100$\,cm$^{-3}$ (Fig.\,\ref{flux_ratios}), $v=40$\,km\,s$^{-1}$ and $n_{arms}=2$, we obtain the value for the flux of matter along the nuclear spirals of $\Phi\approx1.23\times10^{25}$\,g\,s$^{-1}$\,$\times$\,0.004$\approx$\,8$\times10^{-4}$\,M$_{\odot}$\,yr$^{-1}$.

We note that the above calculation is only a rough estimate, as it depends on uncertain parameters, such as the geometry of the flow. 
Under the assumption that the gas moving along the spiral is reaching
the nucleus, we can compare the above calculated rate of ionized 
gas flow to the accretion rate necessary to reproduce the 
luminosity of the LINER nucleus of NGC\,6951, calculated as follows.
The nuclear luminosity can be estimated from the 
H$\alpha$ luminosity of L(H$\alpha$)$=1.6\times\,10^{39}$\,erg\,s$^{-1}$
(from the nuclear flux and reddening quoted by P00) using 
the approximation that the bolometric luminosity is $L_{bol}\approx$
100\,$L$(H$\alpha)=1.6\times\,10^{41}$\,erg\,s$^{-1}$ (Ho 1999; Ho et al. 2001).
This estimate is consistent also with the upper limit on the
X-ray luminosity of this galaxy, $L_X\approx\,4\times\,10^{40}$\,erg\,s$^{-1}$
(Fabbiano et al. 1992; Ho 1999). The mass accretion rate can then
be estimated as:

\begin{equation}
\dot{m}=\frac{L_{bol}}{c^2\,\eta}
\end{equation}
\noindent
where $\eta$ is the efficiency of conversion of the rest mass energy 
of the accreted material into radiation. For a ``standard'' 
geometrically thin, optically thick accretion disk,
$\eta\approx$\,0.1 (e.g. Frank et al. 2002), but for LINERs 
it has been concluded that the accretion disk is geometrically thick,
and optically thin (Ho 2005; Nemmen et al. 2006, Yuan 2007). This
kind of accretion flow is known as RIAF (Radiatively Inefficient 
Accretion Flow; Narayan 2005), and has a typical value for $\eta\approx$\,0.01. 
We use this value to derive an accretion rate of $\dot{m}=1.78\times\,10^{22}$\,g\,s$^{-1}$,
or 2.8$\times\,10^{-4}$\,M${_\odot}$\,yr$^{-1}$. Comparing this
accretion rate $\dot{m}$ with the mass flow of ionized gas along the 
nuclear spirals derived above $\Phi$, we conclude that $\Phi\approx$\,
3\,$\dot{m}$. Thus the flow of ionized gas along the spiral arms 
is of the order of the one necessary to feed the AGN
in NGC\,6951. 

We should nevertheless point out that this
flow is most probably a lower limit to the actual 
inward flow of matter, as we are observing only the ionized
gas, which may be just the ``tip of the iceberg''.
Neutral and molecular gas may also be flowing in,
as suggested by the $\ge$10$^7$\,M$_\odot$ of
molecular gas mass detected at the nucleus by
GB05 and  Krips et al. (2007). 

We now compare the above results obtained for NGC\,6951
with those obtained for NGC\,1097. Storchi-Bergmann (2007)
has calculated the mass flow rate along the nuclear spirals
of NGC\,1097 on the basis of similar IFU observations,
but for a filling factor f=1. If we adopt a more realistic value
of $\sim$ 10$^{-3}$,
it can be concluded that the mass flow rate of ionized
gas is again similar to the accretion rate necessary to
power the LINER nucleus of NGC\,1097. 
Similarly to the case of NGC\,6951, the total mass flow
is probably larger, and could include neutral and molecular gas.
A larger mass flow rate would be consistent with the presence of a
young starburst (age$\sim$10$^6$\,yr) with mass 10$^6$\,M$_{\odot}$ 
discovered in the nuclear HST-STIS UV nuclear spectrum of NGC\,1097 and estimated 
to be closer than 9\,pc from the nucleus (Storchi-Bergmann et al. 2005). 
The gas flowing to the nuclear region is thus not only feeding
the nucleus but has also accumulated enough mass to give
origin to a circumnuclear starburst in the last 10$^6$\,yrs.
The association of recent star-formation with nuclear activity has
been claimed in many recent studies (e.g. Cid Fernandes et al. 2001,
Storchi-Bergmann et al. 2001) and its association in particular 
with dust structures in early-type galaxies has been also 
supported by the recent work of Ferrarese et al. (2006).

\section{Concluding Remarks}

We have presented IFU-GMOS observations of the 
ionized gas in the inner region of the galaxy NGC\,6951 
which evidence the presence of streaming motions
along nuclear spiral arms towards its LINER nucleus.
The streaming motions are observed over an extent
of $\approx$\,200\,pc along two partial spiral arms 
down to $\approx$\,100\,pc from the nucleus.

This is so far only the second galaxy in which such streaming
motions have been mapped, the first being NGC\,1097 
also by our group (Fathi et al. 2006). The relevance of this
result stems from the fact that the presence of dusty nuclear structures 
correlates strongly with nuclear activity (Lopes et al. 2007), 
what suggests that these structures are associated with the actual fuel on its way
in to feed the nuclear SMBH. 

Nevertheless, until recently, there were no
kinematic measurements which showed that these spirals do carry gas inwards.
Such kinematic measurements are difficult to obtain due 
to the fact that the gas kinematics 
of the nuclear region of active galaxies
is usually dominated by outflows from the narrow-line region.
Most kinematic studies do reveal outflows and no inflows
(e.g. Crenshaw \& Kraemer 2007 and references therein). 

In order to observe colder gas that can fall inwards it
is necessary to obtain gas kinematics around  nuclei
with low levels of activity, such that the outflows are small,
if present. This seems to be the case of NGC\,6951:
a LINER nucleus, with a small jet, such that it was
possible to disentangle the outlow related to the jet
from streaming motions along nuclear spirals.
Additional observational constraints to observe
such motions around a galaxy nucleus include 
a favorable orientation of the galaxy to allow the
measurement of the kinematics and strong enough gas emission
to allow such measurements up to
several hundred parsecs from the nucleus, so that
we can constrain the disk rotation and on top of it
measure the residuals from circular motion.

With the present results for NGC\,6951 we
can now say that there are two galaxies with confirmed inward
streaming motions along  nuclear spirals  
-- NGC\,1097 and NGC\,6951 -- giving strong 
support to the hypothesis that these spirals are indeed 
channels through which the matter flows inwards
to feed the SMBH.

The kinematic measurements and an estimated geometry
derived from the observations has allowed us to estimate
the ionized gas mass flow rate along the nuclear spirals
of NGC\,1097 and NGC\,6951. We obtain values which are  
similar to the accretion rate 
necessary to power the AGN.
The total mass flow rate is most probably  
larger, including neutral and molecular gas.
In the case of NGC\,6951, molecular gas 
has been recently observed both at the star forming ring
and inside the ring. In the case of NGC\,1097, the young nuclear 
starburst found by Storchi-Bergmann et al. (2005) supports
also a larger flow of matter to the nucleus in the last 10$^6$\,yrs.
Thus it may be that the inflows
observed along nuclear spiral arms not only feed nuclear SMBH but 
at the same time contribute to the growth of the bulge via star formation.
As a result, both the stellar component of the galaxy and its SMBH at the
nucleus grow after each activity cycle, in agreement
with the scenario implied by the
M$_{BH}\times\sigma$ relation (e.g. Tremaine et al. 2002).

\acknowledgments
We acknowledge the referee for relevant suggestions which have improved the paper.
This work was based on observations obtained at the Gemini Observatory, which is operated 
by the Association of Universities for Research in Astronomy, Inc., under a 
cooperative agreement with the NSF on behalf of the Gemini partnership: the 
National Science Foundation (USA), the Particle Physics and Astronomy Research 
Council (UK), the National Research Council (Canada), CONICYT (Chile), the 
Australian Research Council (Australia), CNPq (Brazil) and CONICET (Argentina).
KF acknowledges support from the Werner-Gren Foundations, the Royal Swedish
Academy of Sciences and project P3/86 from the Instituto de Astrof\'isica de 
Canarias. This work is also based on observations with the NASA/ESA Hubble Space Telescope 
obtained at the Space Telescope Science Institute, which is operated by 
the Association of Universities for Research in Astronomy, Incorporated, under 
NASA contract NAS5-26555. TSB, OLD and RAR acknowledge support from the Brazilian
Institutions CNPq and CAPES.

\end{document}